\documentclass[prd,onecolumn,preprint,nofootinbib]{revtex4}
\usepackage{graphicx}

\usepackage[plainpages=false, colorlinks=true, anchorcolor=blue, linkcolor=blue, citecolor=blue, bookmarks=false]{hyperref}
\usepackage{color}
\usepackage{float}
\usepackage{amsfonts,amsmath,amssymb}
\usepackage{natbib}
\usepackage{array}
\begin{document}
\newcolumntype{P}[1]{>{\centering\arraybackslash}p{#1}}
\pdfoutput=1
\newcommand{\jcap}{JCAP}
\newcommand{\araa}{Annual Review of Astron. and Astrophys.}
\newcommand{\rthis}[1]{\textcolor{black}{#1}}
\newcommand{\aj}{Astron. J. }
\newcommand{\mnras}{MNRAS}
\newcommand{\apjl}{Astrophys. J. Lett.}
\newcommand{\apjs}{Astrophys. J. Suppl. Ser.}
\newcommand{\aap}{Astron. \& Astrophys.}
\renewcommand{\arraystretch}{2.5}
\title{Search for spatial coincidence between IceCube neutrinos and  gamma-ray bright red dwarfs}
\author{Fathima \surname{Shifa M}}
\altaffiliation{E-mail:ph23mscst11012@iith.ac.in}

\author{Shantanu \surname{Desai}}
\altaffiliation{E-mail: shntn05@gmail.com}

\begin{abstract}
We search for a spatial coincidence between  high energy neutrinos detected by the IceCube neutrino detector and  ten   red dwarfs  which have been observed in gamma-rays. For our analysis, we use the 
unbinned maximum likelihood method to look for a statistically  significant excess. We do not find any such spatial association between any of  the red dwarfs and  IceCube-detected neutrinos. Therefore, we conclude that none of the gamma-ray bright  red dwarfs contribute to the diffuse neutrino flux measured by IceCube.    
\end{abstract}

\affiliation{Department  of Physics, IIT Hyderabad,  Kandi, Telangana-502284, India}

\maketitle
\section{Introduction}
\label{sec:intro}
The origin of majority of the IceCube  diffuse neutrino flux in the TeV-PeV energy  range~\cite{IceCubescience} is still  unknown~\cite{Halzen23}. Knowing the sources of the diffuse neutrino background flux will help us unravel a large number of problems in Physics and Astronomy from origin of ultra-high energy cosmic rays to beyond the standard model of particle physics~\cite{Halzen24}. Although, searches by the IceCube collaboration have shown evidence for neutrino emission from some  point sources such as  NGC 1068, TXS 0506+056, NGC 4151, and PKS 1424+240, the majority of IceCube neutrinos cannot be attributed to any astrophysical sources~\cite{IceCubedata}. 
 Searches for spatial coincidence with a plethora of extragalactic  sources, which could contribute to the diffuse neutrino flux have been done. A non-exhaustive list  of such extra-galactic sources  includes  AGNs and GRBs~\cite{Kamionkowski,Hooper,LuoZhang,Smith21,Li22,Icecube10,IceCubeGRB,IceCubeAGN,IceCubeblazars}, FRBs~\cite{Zhang21,Desai23,IceCubeFRB}, high energetic events from the Fermi-LAT catalog~\cite{Li22}, merging galaxies~\cite{Laha}, and  galaxy catalogs using the 2MASS survey~\cite{IceCube2MASS} and WISE-2MASS sample~\cite{Fang24}. 

About a year ago,  the IceCube collaboration found 4.5$\sigma$ evidence for neutrino emission from the Galactic plane using cascade events~\cite{Science}. Although the observed signal is consistent with diffuse emission from the  galactic plane, some of the signal could arise from a population of unresolved point sources such as supernova remnants, pulsar wind nebulae or unidentified TeV Galactic sources~\cite{Science}.

In a similar vein, a galactic contribution to the all-sky  IceCube diffuse neutrino flux cannot be excluded and has been estimated to be about 10-20\%~\cite{Palladino,Troitsky}. 
Therefore, searches for coincidences with multiple galactic sources such as supernova remnants, X-ray binaries, pulsars, pulsar wind nebulae, LHAASO and HAWC sources,  open clusters have also been carried out~\cite{Lunardini,Marfatia,IceCubeXRB,IceCubePWN,Kovalev22,LHAASO,Li22,Kumar,Pasumarti,Pasumarti2}. A large number of galactic sources have also been targeted for neutrino astronomy by the previous generation of underground neutrino detectors~\cite{MACRO,sknuastro,skshowering,Desai22} 

In this work, we look for high energy neutrinos from red dwarfs. Red dwarfs (sometimes also referred to as M-dwarfs) are one of the smallest stars on the main sequence with masses between 0.075-0.5 $M_{\odot}$, having surface temperatures between 2,500-5,000 K~\cite{Edgeworth}, and frequently emit flares with energies between $10^{32}$ and $10^{35}$ ergs with a power-law index between 1.7-2.4~\cite{Yang17}. Energetic outbursts have been detected from red dwarfs at high energies from hard X-rays to TeV gamma-rays~\cite{Drake,Song,Shalon}. Most recently, the SHALON TeV gamma-ray telescope  detected very high energy gamma-ray emission  between 800 GeV and 20 TeV from eight red dwarfs, namely, 
V388 Cas, V547 Cas, V780 Tau, V962 Tau, V1589 Cyg, GJ 1078, GJ 3684 and GL 851.1~\cite{Shalon}. Subsequently, these red dwarfs have been proposed as sources of ultra high energy cosmic rays~\cite{rdcosmicrays,Sinit}. 
However, no corresponding gamma-ray emission was seen at GeV~\cite{Huang24} or MeV energies~\cite{Niharika}.~\citet{Song} reported a $5\sigma$ detection of pulsed GeV $\gamma$-ray emission from another M-dwarf TVLM 513-46546 with  Fermi-LAT.
A very hard X-ray outburst was detected by the SWIFT-BAT telescope from another M-dwarf, viz 
DG CVn from 0.3-100 keV in April 2014~\cite{Drake}, although no coincident gamma-ray emission was seen from Fermi-LAT~\cite{Loh}. Although there are no estimates  for high energy neutrino  flux emission from  red dwarfs, these red dwarfs could generate  cosmic rays with energies  up to $10^{14}$ eV in their flaring states~\cite{Sinit}. TeV gamma-rays detected from these red dwarfs indicate that these red dwarfs produce flares with total energies of $10^{32}-10^{35}$ ergs~\cite{rdcosmicrays,Sinit}.  If one considers the statistics of M-dwarfs in our galaxy and assuming the frequency of flares to be around 36/year for about 10\% of the observed  red dwarfs, the total energy released from these flares matches the total energy of cosmic rays in the galactic disk of $\sim 10^{54}$ ergs~\cite{Sinit19}.
During these flares the accelerated protons with energies up to $10^{14}$ eV undergo spallation with the atmosphere outside the star~\cite{Sinit}. Note that  similar mechanisms for production of TeV gammma-rays have also been  proposed for the Sun~\cite{Zhou17}. As argued in ~\cite{Sinit}, these interactions produce pions. Consequently, one would also expect neutrino emission from the aforementioned hadronic interactions along with gamma-rays produced from pion decay. Therefore, these considerations provide enough motivation to search for neutrino emissions  to complement recent searches for MeV~\cite{Niharika} and GeV~\cite{Huang24} gamma-rays from these red dwarfs.

Therefore, in this work we look for neutrino emission from the aforementioned 10 red dwarfs using the publicly available muon track data from IceCube. We have previously used this data to look  for neutrino emission from pulsars and FRBs~\cite{Desai23,Pasumarti,Pasumarti2}. The outline of this manuscript is as follows. The  neutrino dataset  used for the  analysis is discussed in Section~\ref{sec:dataset}. The analysis and results are discussed in Section~\ref{sec:results}.  We conclude in Section~\ref{sec:conclusions}.

\section{Dataset}
\label{sec:dataset}
The neutrinos used in this work are taken from the IceCube 10 year muon track data~\cite{IceCubedata}. IceCube is a detector located at the South Pole, \rthis{which detects neutrinos through the Cherenkov radiation emitted by the charged leptons traveling through the ice.}

This dataset  consists of  1,134,431 \rthis{track-like neutrino candidates}, which were  collected  between  April 2008 (IC-40) and July 2018 (IC 86-V11) from four different phases of the experiment, each having different livetime.  Note that for our analysis we have used the augmented dataset analyzed in ~\cite{Beacom}\footnote{This dataset is available at \url{https://github.com/beizhouphys/IceCube_data_2008--2018_double_counting_corrected}}, which corrects for 38 duplicate events found in the IceCube public muon track data due to angular  reconstruction error~\cite{Beacom}. For each neutrino, the dataset consists of  right ascension (RA), declination ($\delta$),  error in the position,  and reconstructed muon energy. 
\section{Analysis and Results}
\label{sec:results}
For our analysis, we are using the unbinned maximum likelihood ratio method (See ~\cite{Pasumarti} and references therein, and first proposed in ~\cite{Montaruli}). We use neutrinos within a declination of $\pm 5^{\circ}$ of the red dwarf. For a dataset containing $N$ events, if $n_s$ is the number of signals attributed to the red dwarf, then the likelihood of the entire dataset is given by:
\begin{equation}
\mathcal{L} (n_s) = \prod_{i=1}^N \left[\frac{n_s}{N} S_i + (1-\frac{n_s}{N}) B_i\right]
\label{eq1}
\end{equation}
where $S_i$ is the signal PDF and $B_i$ is the background PDF. Here the signal PDF is given by:
\begin{equation}
S_i = \frac{1}{2\pi\sigma_i^2}e^{-(|\theta_i-\theta_s|)^2/2\sigma_i^2}
\label{eq:2}
\end{equation}
where $|\theta_i-\theta_s|$ is the angular distance between the  neutrino and the red dwarf, $\sigma_i$ is the angular uncertainty in the neutrino position, in radians.The background PDF can be obtained from  the solid angle within the declination ($\delta$) of  $\pm 5^{\circ}$ around each red dwarf and is given as follows~\cite{Montaruli}:
\begin{equation}
B_i=\frac{1}{\Omega_{\delta \pm 5^{\circ}}}
\end{equation}
\rthis{We then evaluate $\mathcal{L} (n_s)$ for different values of $n_s$, and designate the value of $n_s$,  for which this likelihood is maximized as   $\hat{n}_s$.}
Now to ascertain the significance of the signal, we define the  test statistics ($TS$) \rthis{based on the   likelihood ratio, with respect to the null hypothesis in the absence of a signal, ($\mathcal{L} (0)$}) as follows:
\begin{equation}
TS  = 2 \log \frac{\mathcal{L} (\hat{n}_s)}{\mathcal{L} (0)}.
\label{eq:ts}
\end{equation}
For the null hypothesis, $TS$ behaves like a $\chi^2$ distribution for one degree of freedom~\cite{Wilks}. 
The detection significance  can be obtained from  $\sqrt{TS}$. This statistics is also used in gamma-ray astronomy (eg.~\cite{Manna}). 
For a statistically significant detection of $>5\sigma$, TS $> 25$.
For each of our red dwarfs, we determine  $\hat{n}_s$ \rthis{and evaluate $TS$ according to Eq~\ref{eq:ts}.   These values of TS, along with the  values of $\hat{n}_s$ can be found in Table~\ref{table1}}. As we can see, none of the red dwarfs show a TS value $> 25$. The largest TS value is seen for GJ 1078 with TS of 2.38, whose significance is  less than $2\sigma$. Therefore, the observed signal events are consistent with background, and there is no evidence for any spatial association between red dwarfs and IceCube neutrinos, which are observed  as muon tracks.  Consequently, we  calculate the 95\% c.l. upper flux limits by calculating the value of $n_s$ for which $\Delta (TS)=-3.84$~\cite{PDG}, where $\Delta TS=TS-TS_{max}$.
\section{Conclusions}
\label{sec:conclusions}
In this work, we searched for  a spatial coincidence between  IceCube neutrinos and 10 gamma-ray bright red dwarfs.   The neutrinos used in this work are taken from the publicly released 10 year muon track data observed between 2008-2018, and consisting of  113443 neutrinos. Eight of the 10 red dwarfs were detected at TeV gamma rays by the SHALON telescope. Two others were detected in gamma-rays and X-rays from Fermi-LAT and SWIFT.  We analyzed these data using the unbinned maximum likelihood method. A tabular summary of our results  for all the  red dwarfs can be found  in Table~\ref{table1}. All the red dwarfs show a test statistic value of less than two,\rthis{ with significance less than $2\sigma$.}  Therefore,  we conclude that none of the above red dwarfs contribute to the Ice Cube diffuse neutrino flux.
\begin{center}
\begin{table}[ht]

    \begin{tabular}{|l|l|l|l|l|l|l|}
    \hline
    \centering
        \textbf{Red dwarf} & \textbf{RA ($^{\circ}$)} & \textbf{$\delta (^{\circ}$)} & $\hat{n}_s$  & $TS$ & Upper limit \\ \hline
        V388 Cas & 15.8326 & +62.365 &  0.0  & 0.0 &16.2 \\ \hline
        V547 Cas & 8.1229 & +67.235 &0.0 &   0.0  &  16.7\\ \hline
        V780 Tau & 85.107 & +24.802 &6.00 & 0.35 & 29.8\\ \hline
        V962 Tau & 86.466 & +22.879 &0.0 &    0.0 &  13.8 \\ \hline
        V1589 Cyg & 310.705 & +41.38 &0.0 &    0.0  &  12.7\\ \hline
        GJ 1078 & 80.95 & +22.54 & 20.39 &  2.38 & 51.2\\ \hline
        GJ 3684 & 176.77 & +70.03 & 0.0 &   0.0  &  14.9\\ \hline
        GL 851.1 & 333.0267 & +31.56 & 0.0 &    0.0  & 11.1\\ \hline
        TVLM 513-46546&225.284&22.834&0.0 &0.0 &20.8 \\ \hline
        DG CVn&202.944&29.28&2.42&  0.05 & 27.2\\  \hline
        
    \end{tabular}
 \caption{\label{table1}Results from spatial coincidence analysis between IceCube neutrinos and red dwarfs. None of the red dwarfs show a detection significance $> 5\sigma$, corresponding to TS$>25$. Therefore, we do not find any spatial association between the red dwarfs and IceCube neutrinos. The last column shows the 95\% c.l. upper limit on the number of signal neutrino events.}
\end{table}
\end{center}

\section*{Acknowledgments}
We are grateful to Vibhavasu Pasumarti for making the analysis codes used in ~\cite{Pasumarti} publicly available and the anonymous referee \rthis{for constructive feedback on the manuscript.}

\bibliography{main}
\end{document}